# Optical properties of mist CVD grown α-Ga$_2$O$_3$

Usman Ul Muazzam[1], Prasad chavan[1], Srinivasan Raghavan[1], R. Muralidharan[1], Digbijoy N Nath[1]

[1] *Centre for Nano Science and Engineering, Indian Institute of Science, 560012, Bangalore, India*

We report on the study of optical properties of mist CVD grown α-Ga$_2$O$_3$ with the observation of excitonic absorption in spectral responsivity measurements. 163 nm of Ga$_2$O$_3$ was grown on sapphire using Ga-(acac)$_3$ as the starting solution at a substrate temperature of 450°C. The film was found to be crystalline and of α-phase with an on-axis full width at half maximum (FWHM) of 92 arcsec as confirmed from X-ray diffraction scans. The Tauc's plot extracted from absorption spectroscopy exhibited two transitions in the UV regime at 5.3 eV and 5.6 eV, corresponding to excitonic absorption and direct band-to-band transition respectively. The binding energy of exciton was extracted to be 114 meV from spectral responsivity (S.R) measurements. Further, metal-semiconductor-metal (MSM) photodetectors (PD) with lateral inter-digitated geometry were fabricated on the film. A sharp band edge was observed at 230 nm (~ 5.6 eV) in the spectral response with peak responsivity of ~1 A/W at a bias of 20 V. The UV to visible rejection ratio was found to be ~ 100 while the dark current was measured to be ~ 0.1 nA.

The various polymorphs of Ga$_2$O$_3$, with their wide band gap of 4.6-5.3 eV eV[1,2], have attracted attention of the device community for their promises in the areas of high-power switching[3], deep-UV optoelectronics[4], gas sensors[5], high-temperature and transparent electronics[6]. UV-C photodetectors for instance, are useful in UV astronomy, bio-medical and forensic applications, and for missile plume detection in the strategic sector[7–9]. Although the β phase is the most stable among the five different polymorphs (α, β, γ, ε, δ) of Ga$_2$O$_3$ and thus has been the most widely investigated, there has been an increasing interest in α-Ga$_2$O$_3$ in recent times. It has a corundum crystal structure and has been predicted to have the highest bandgap (~ 5.3 eV[10]) among all the polymorphs of Ga$_2$O$_3$. This makes α-Ga$_2$O$_3$ an attractive candidate for ultra-high breakdown transistors and deep-UV opto-electronics at sub-240 nm wavelengths. The growth of α-Ga$_2$O$_3$, which requires relatively low temperatures (430°C – 470°C) [10], has been reported using approaches such as atomic layer deposition (ALD), mist chemical vapor deposition and molecular beam epitaxy[11,12]. Although there is a report on the demonstration of field effect transistor (FET) based on mist CVD grown α- Ga$_2$O$_3$[13] , the investigation of the growth as well as structural, optical and electrical transport properties of this emerging polymorph of Ga$_2$O$_3$ is still at an embryonic stage. In this letter, we report on the study of optical – in particular excitonic - properties of mist CVD grown α-Ga$_2$O$_3$ with a subsequent realization of a high-responsivity solar blind UV-C photodetectors.

The mist CVD system used for the growths has been developed in-house and consists of two parts, the reactor and the mist generator. A volume of 0.33 mole 5N pure gallium acetylacetonate Ga(acac)$_3$ dissolved in de-ionised water (DI) was used as the source of gallium precursor. Small amount (0.1 ml) of HCl was added to ensure the complete dissolution of Ga(acac)$_3$ in DI water. This solution was then ultra-sonicated at a frequency of 1.6 MHz using the mist-generator. The generated mist was directed to the deposition zone using N$_2$ (500 sccm) as carrier gas. c-plane sapphire wafer of 2-inch diameter was diced into 1 cm x 1 cm pieces and were solvent cleaned in acetone, isopropyl alcohol and rinsed with DI water. For each growth run, a piece of sapphire was placed inside the deposition zone using a quartz tube with diameter of 40 mm. The growth was carried out for one hour at a temperature of 450 °C and at atmospheric pressure.

The XRD scans were carried out using a Rigaku SmartLab, Cu-Ka radiation X-ray diffraction system. The film grown on sapphire was confirmed to be α-Ga$_2$O$_3$ from θ-2θ scan. Figure1 shows (0006) reflection of α-Ga$_2$O$_3$, and the inset to figure1 shows the symmetric rocking curve with an FWHM of 92.2 arcsec indicative of a low screw dislocation density in the epi-layer. The surface morphology of the as-deposited film was studied using atomic force microscope (Dimension ICON, Bruker) and the rms roughness was found to be 2 nm as shown in Figure 2(a), which confirms the smoothness of the film. The film thickness was found to be 163 nm from ellipsometry measurements. Figure 2(b) shows the image of the as-grown film as obtained from scanning electron microscope (GEMINI Ultra 55, FE-SEM, Carl Zeiss), indicating that the layer is continuous and uniform.

a) Corresponding author email: usmaanm@iisc.ac.in
digbijoy@iisc.ac.in

Absorption measurement was done using UV-Vis setup (UV-3600, UV-VIS-NIR spectrophotometer, Shimadzu). The Tauc's plot (figure 3) exhibited a distinct kink in addition to the primary absorption edge. The first edge at 5.3 eV corresponds to excitonic transition while the kink with sharp transition corresponds to band-to-band absorption at 5.6 eV when extrapolated linearly to intersect the x-axis and indicates the band gap of α-$Ga_2O_3$.

Photodetectors with metal semiconductor metal (MSM) layout in an interdigitated geometry were fabricated on the as-grown α-$Ga_2O_3$ sample using standard i-line lithography process. The device schematic is shown in Fig. 4(a). Ni (20nm)/Au (100nm) stack was deposited using sputtering to form Schottky contact. Each MSM detector as shown in Fig. 4(b), comprised of seventeen pairs of interdigitated fingers where each finger had a width of 4 μm and the finger spacing was 6 μm. The active area of each device was 260 μm x 300 μm.

Spectral responsivity (SR) measurement was done using a quantum efficiency setup, the details of which are reported elsewhere[14]. The SR spectra exhibited a primary peak at 230 nm corresponding to band-to-band absorption while an excitonic peak could be observed at 235 nm. The binding energy of exciton, estimated from the difference between the two peaks, was found to be 114 meV, which is in close agreement with earlier reported values[15]. This is also the first report of observation of excitonic peak in spectral response of any polymorph of gallium oxide.

Raman spectra was recorded in the backscattering geometry using 532 nm laser, the light was then collected using 100x objective in backscattered geometry and analysed using LabRAM HR, Horiba spectrometer. Figure 5. shows Raman spectra of α-$Ga_2O_3$. The corundum structure of α-$Ga_2O_3$ belongs to -3m ($D_{3d}$) point group and R-3c ($D_{3d}^6$) space group. According to group theory analysis the irreducible representation of zone-centre optical mode is:

$$\Gamma = 2A_{1g} + 2A_{1u} + 3A_{2g} + 2A_{2u} + 5E_g + 4E_u \qquad (1)$$

In addition, unit cell of α-$Ga_2O_3$ is centrosymmetric thus all vibrations that are Raman allowed are infrared forbidden and vice-versa. The $A_{1g}$ and $E_g$ are Raman active, $A_{2u}$ and $E_u$ are infrared active, and $A_{1u}$ and $A_{2g}$ vibrations are neither Raman nor infrared active. The spectrum shows $A_{1g}$(LO) phonon mode at 216 cm$^{-1}$. This mode is attributed to Ga atoms vibrating against each other along c-axis. The high frequency $E_g$ mode at 430 cm$^{-1}$ is due to lighter O atom vibrations perpendicular to c-axis[16]. Low intensity of Raman modes of α-$Ga_2O_3$ may be due to thinner sample.

In most of the oxide semiconductors, the excitonic binding energy is larger than the Rydberg exciton effective energy which is given by:

$$E_{exo} = R_y \frac{\frac{\mu}{m_o}}{\left(\frac{\epsilon_s}{\epsilon_o}\right)^2} \qquad (2)$$

$$\frac{1}{\mu} = \frac{1}{m_e} + \frac{1}{m_h}$$

Here, $R_y$ is the Rydberg energy which has value of 13.6 eV, $\mu$ is the reduced mass of exciton, $m_o$ is free electron mass, $\epsilon_s$ the static dielectric constant of $Ga_2O_3$ which is 10[17]. Since $m_h \gg m_e$, $\mu$ is approximately taken as $m_e$ which is $0.276 m_o$[18]. From equation (1), the excitonic binding energy is found to be around 37.53 meV which is underestimated because we have not considered the interaction between LO phonons and excitons. Polar materials have more than one atom per unit cell having non-zero Born effective charge; thus, atomic displacement corresponding to polar LO phonons can give rise to microscopic Born effective electric fields at long wavelengths leading to strong coupling between LO phonons and excitons[19]. Interaction between polar optical phonons and excitons can be described using Frohlich coupling constant, which is given by[20]:

$$\alpha_F = \frac{q^2}{8\pi\epsilon_o \hbar} \sqrt{\frac{2m_c}{\hbar\omega_o}} \left(\frac{1}{\epsilon_\infty} - \frac{1}{\epsilon_s}\right) \qquad (3)$$

Using equations (2) and (3) $E_{ex}$ can be estimated to be 91.71 meV.

This value is very close to the value estimated from spectral responsivity. To estimate the dissociation field for exciton, polaron (coupling of LO phonon with electron) radius was calculated using[21]:

$$a_p = \sqrt{\frac{2\hbar}{m\omega_{LO}}} \qquad (4)$$

a)  Corresponding author email: usmaanm@iisc.ac.in
    digbijoy@iisc.ac.in

where $a_p$ is Polaron radius. From equation (4), $a_p$ was found to be 32.2 Å, which corresponded to a dissociation field of 0.354 MV/cm.

Figure6 (a) shows the variation of responsivity with wavelength at different voltages on linear scale (5 V, 10 V, 15 and 20 V). Inset to figure 6(a) shows the same in log scale. The peak responsivity was measured to be 0.95 A/W at 230 nm at a bias of 20 V. The UV to visible rejection ratio was calculated by dividing the responsivity value at 230 nm by that at 400 nm, and was found to be$> 10^2$ at 20 V.

Figure 6(b) shows the current-voltage (I-V) characteristic of the detectors under dark and under illumination at 230 nm. The photo current was found to be 85 nA while the dark current was measured to be 137 pA at an applied bias of 20 V, indicating a photo-to-dark current ratio exceeding two orders of magnitude.

Figure 6(c) shows the variation of peak responsivity with applied voltage at an illumination of 230 nm. The peak responsivity was found to increase with an increase in applied voltage.

The theoretical value of responsivity at 230 nm, assuming a quantum efficiency of 100%, is 185 mA/W. This is much smaller than the peak responsivity value of 518 mA/W at 5 V (at 230 nm) measured in this work, even at a relatively low bias of 5 V, implying that there is gain in the devices[22–24]. This gain could be because of oxygen vacancies[25], which act as hole trapping centres in the bulk of the semiconductor which could leads to photoinduced barrier lowering[22] resulting in an increase in transit time.

In conclusion, we have reported on the study of growth and photo-response properties of mist CVD grown α-$Ga_2O_3$ on c-plane sapphire. Solar blind deep-UV photodetectors realized on these samples exhibited high responsivity of 0.5 A/W at 5 V bias with a sharp peak at 5.5 eV, low dark current of ~ pA and UV-to-visible rejection ratio exceeding two orders of magnitude. We reported the first observation of excitonic peak in spectral responsivity with an excitonic binding energy of 114 meV. This work is expected to aid further in the understanding of optical properties of α-$Ga_2O_3$ Towards realizing high-performance deep-UV optoelectronics based on gallium oxide.

This work was supported in part by the Ministry of Electronics and Information Technology (MeitY), and in part by the DST through the NNetRA.

a) Corresponding author email: usmaanm@iisc.ac.in
digbijoy@iisc.ac.in

**Figures**

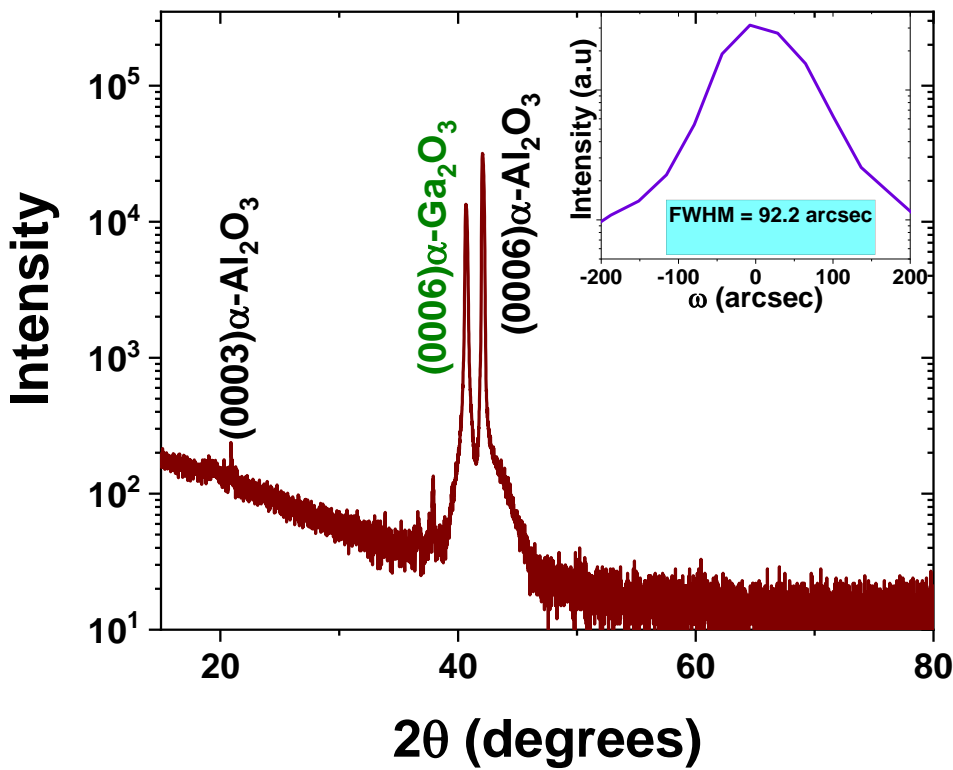

**Figure: 1.** XRD θ-2θ diffraction pattern of as deposited film of α-Ga$_2$O$_3$, inset shows rocking curve plot of (0006) peak of α-Ga$_2$o$_3$ with FWHM of 92.2 arcsec.

a) Corresponding author email: usmaanm@iisc.ac.in
digbijoy@iisc.ac.in

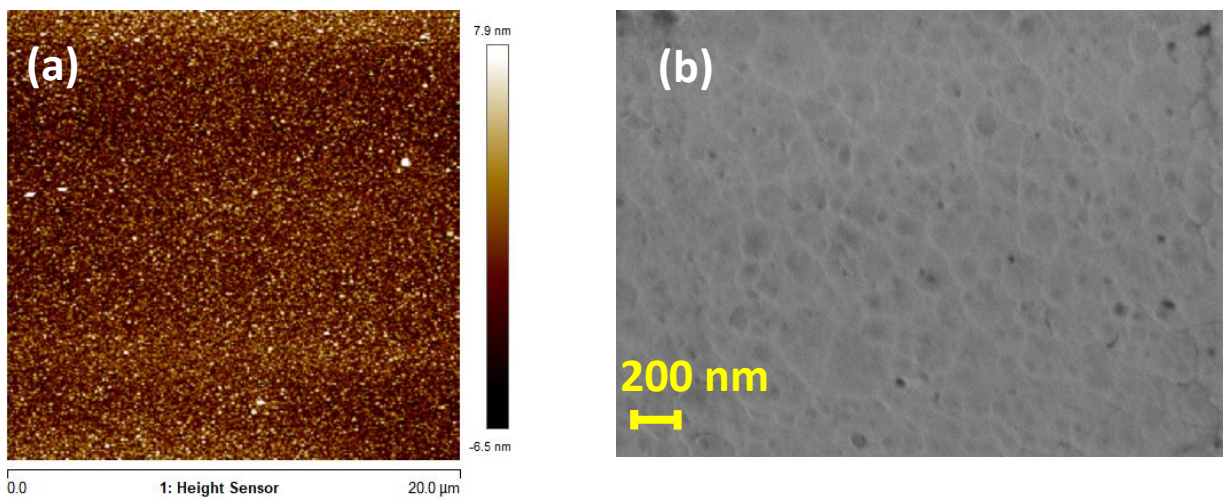

**Figure2. (a)** AFM scan image showing R.M.S roughness of 2 nm. (b) SEM micrograph showing smooth morphology of as deposited film.

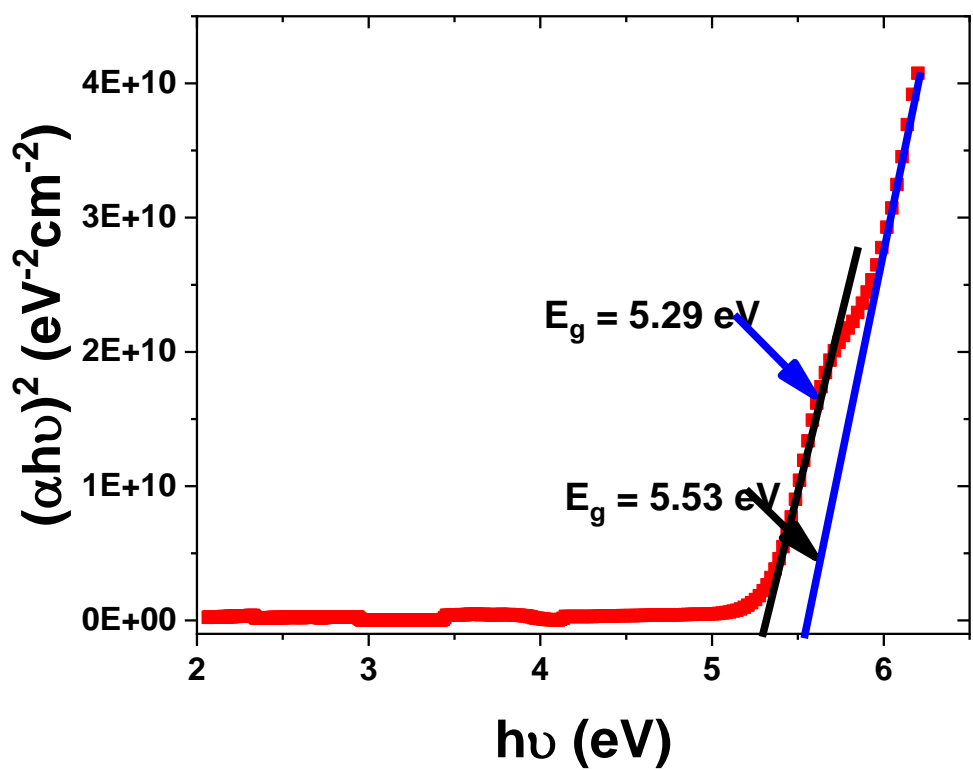

**Figure3.** Tauc's plot showing two transitions.

a) Corresponding author email: usmaanm@iisc.ac.in
digbijoy@iisc.ac.in

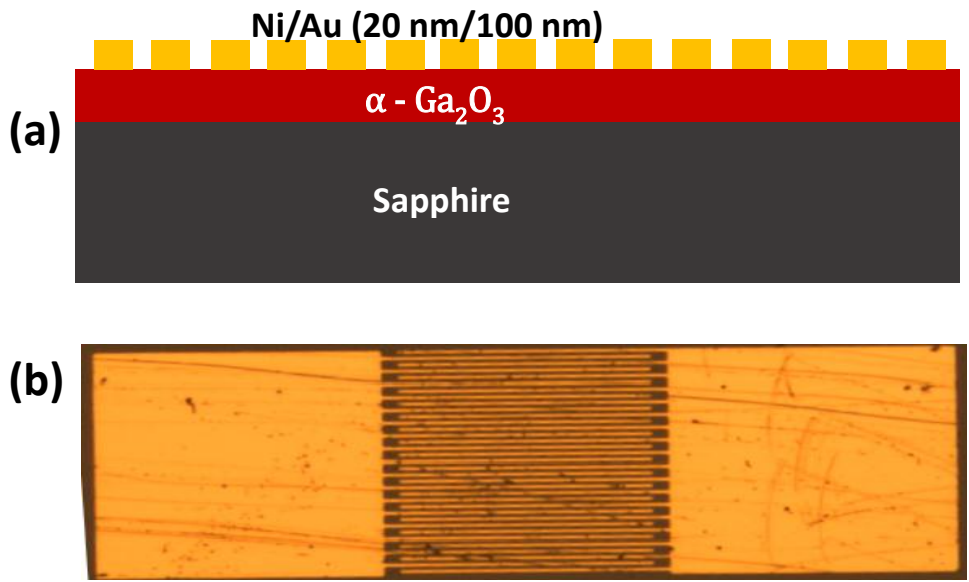

**Figure4** (a) Schematic of MSM photodetector (side view). (b) Optical micrograph of MSM photodetector (top view).

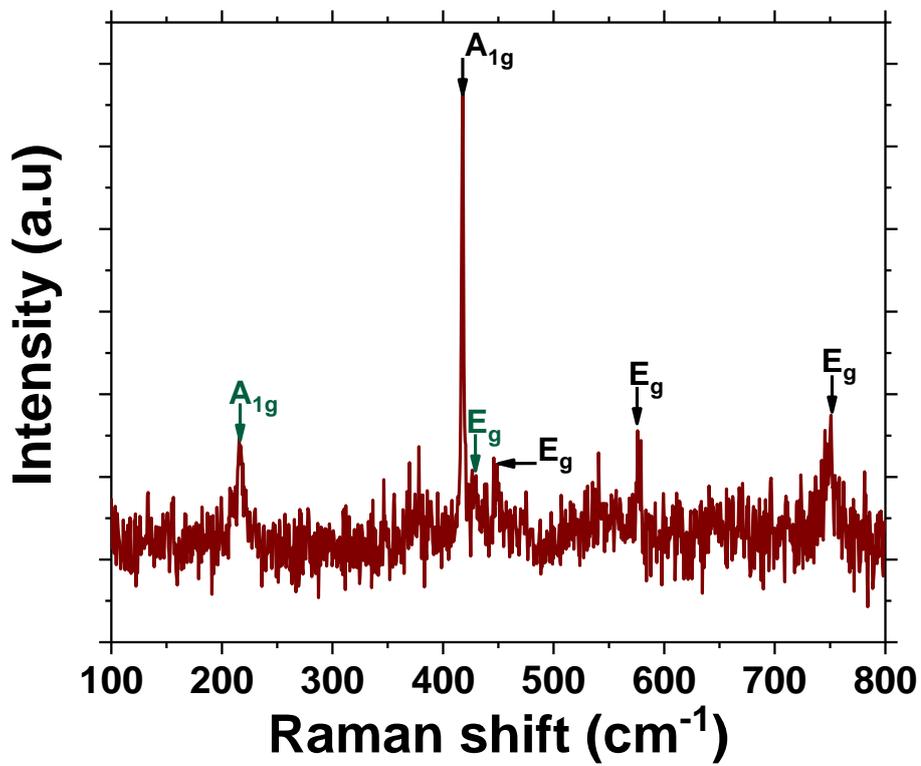

**Figure5** (a). Raman spectra of as deposited film. The green and black labels correspond to α-Ga$_2$O$_3$ and Sapphire peaks respectively.

a) Corresponding author email: usmaanm@iisc.ac.in
digbijoy@iisc.ac.in

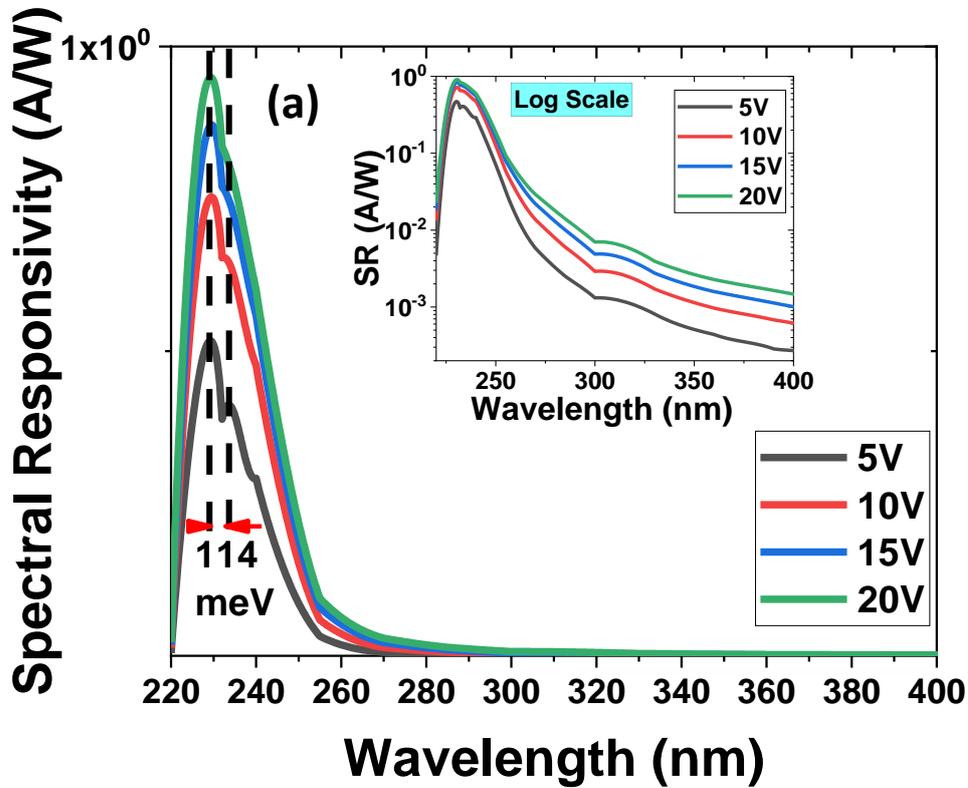

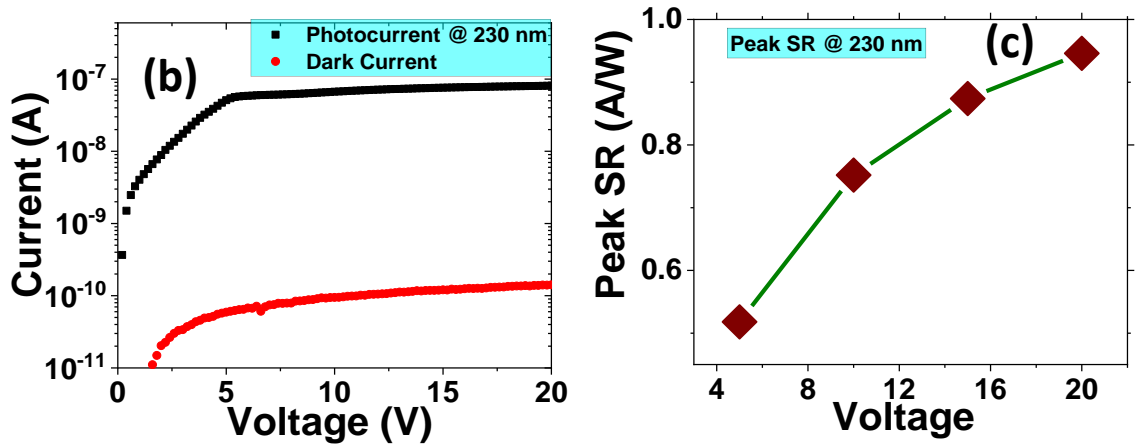

**Figure6** (a). Shows variation of spectral response with wavelength as a function of voltage in linear scale, inset shows variation of SR with wavelength as a function of voltage in log scale. Also can be seen U.V-Visible rejection ratio is > $10^2$. (b) Variation of photocurrent (at 230 nm) and dark current with applied voltage. (c) Variation of peak SR (at 230 nm) with applied voltage.

a)   Corresponding author email: usmaanm@iisc.ac.in
                               digbijoy@iisc.ac.in